\documentclass[aps,prb,twocolumn,groupedaddress]{revtex4-1}
\usepackage[dvips]{graphicx}
\usepackage{amsmath}
\usepackage{amssymb}
\usepackage{color}
\usepackage{bm}

\newcommand{\ket}[1]{|#1\rangle}
\newcommand{\bra}[1]{\langle#1|}
\newcommand{\brket}[3]{\langle#1|#2|#3\rangle}
\newcommand{\outpr}[2]{\ket{#1}\bra{#2}}
\newcommand{\overlap}[2]{\langle#1|#2\rangle}
\newcommand{\avg}[1]{\langle#1\rangle}
\newcommand{\tm}[1]{\textrm{#1}}
\def\dbar{{\mathchar'26\mkern-12mu d}}

\begin{document}


\title{Theory of Entropy Production in Quantum Many-Body Systems}


\author{E. Solano-Carrillo}
\author{A. J. Millis}
\affiliation{Department of Physics, Columbia University, New York, NY 10027, USA}


\begin{abstract}
We define the entropy operator as the negative of the logarithm of the density matrix, give a prescription for extracting its thermodynamically measurable part, and discuss its dynamics. For an isolated system we derive the first, second and third laws of thermodynamics. For weakly-coupled subsystems of an isolated system, an expression for the long time limit of the expectation value of the rate of change of the thermodynamically measurable part of the entropy operator is derived and interpreted in terms of entropy production and entropy transport terms. The interpretation is justified by comparison to the known expression for the entropy production in an aged classical Markovian system with Gaussian fluctuations and by a calculation of the current-induced  entropy production in a conductor with electron-phonon scattering.\end{abstract}

\pacs{}

\maketitle
\section{Introduction}
Attempts to show how nonequilibrium thermodynamic behavior emerges from the underlying quantum mechanics of individual particles is now being dubbed quantum thermodynamics. \cite{Gemmer,Millen,Goold,Vinjanampathya} Several approaches have arisen, revealing important aspects in this endeavor, such as how thermal fluctuations and external driving mechanisms affect the stochastic course of nonequilibrium processes of small systems, \cite{Bustamante} which has led to fluctuation theorems \cite{Esposito,Campisi,Dorner} going beyond the results from the Kubo linear response theory, as well as generalized fluctuation-dissipation relations as studied in isolated quantum systems after a quench. \cite{Khatami,Marcuzzi,Essler} Other aspects, more in the spirit of traditional nonequilibrium statistical mechanics, \cite{Penrose} include thermalization in isolated quantum systems, \cite{Polkovnikov,Yukalov,DAlessio,GogolinE,EisertFG} and the establishment of steady states in open quantum systems. \cite{Dutt,Esposito2,Hsiang,Yuge,Ness,Manzano,Xu} A unified treatment along the lines of the classical theory of nonequilibrium thermodynamics is of crucial importance for a clear identification of the quantum-to-classical correspondence and the new features brought about by fully quantum-mechanical nonequilibrium behavior.

The remarkable success of the classical theory \cite{Prigogine,Groot,Jou2,Jou} in the description of macroscopic phenomena in fluids motivates us to ask what are the basic ingredients of this formalism that such a unified treatment of quantum thermodynamics must also contain. We remind that the building blocks of the classical theory are: (i) macroscopic observables,  explicitly defined as a set of thermodynamically measurable or slowly-varying quantities, (ii) conservation laws for these variables and, as a foundational pillar, (iii) an entropy balance equation is established, splitting the rate of change of entropy as a part which is irreversibly produced, in accordance with the second law of thermodynamics, and a part which is transported. The validity of this theory relies on the \emph{local equilibrium} assumption, whereby the nonequilibrium thermodynamic entropy is considered locally as a function of the same extensive variables as in equilibrium. 

Although significant attempts to give a meaning to entropy out of equilibrium \cite{Lieb} have long been known in quantum statistics, \cite{Mori1,Mori2,Zubarev,Zubarev2,Robertson} a complete theory of quantum entropy production has not been provided yet. The main problem is how to conceive an adequate quantum entropy balance equation without assuming local equilibrium.

For an isolated system, there is no entropy to be transported outside the system and hence the entropy balance equation reduces to finding the right quantum expression for entropy whose rate of change is non-negative, according to the second law of thermodynamics, this rate then being the entropy production. Important efforts have been devoted to obtain such an expression from the density matrix, \cite{Penrose,Polkovnikov2,Ikeda} but the third law of thermodynamics, involving the vanishing entropy of pure states has not been satisfactorily established. 

On the other hand, for a subsystem of an isolated system the establishment of a quantum entropy balance equation has been partially addressed \cite{Spohn2,Spohn,EspositoNJP,Hossein,Mehta,Suzuki,Suzuki2} by assuming that the rate of change of an adopted expression for the nonequilibrium entropy of the subsystem, obtained from the reduced density matrix, is directly connected, as in the classical theory, with the rate of change of its energy. This involves the identification of a microscopic expression for heat which is not unique \cite{Pucci} and therefore quite problematic, but most importantly, does not constitute a full deviation from the local equilibrium assumption, as we show later.

The purpose of this paper is to provide a more general treatment of quantum entropy production and then lay the foundation of a unified theory of quantum thermodynamics in close correspondence with the classical theory. We introduce a \emph{new} thermodynamic entropy operator $\hat{\mathcal{S}}_t$ for isolated quantum many-body systems and show that the rate of change of its expectation value is non-negative, according to the second law of thermodynamics. Unlike previous approaches, we establish the third law of thermodynamics as a well-defined vanishing of the thermodynamic entropy for pure states. 

The quantum entropy balance equation for a given subsystem of an isolated system is obtained by first studying the time evolution of $\avg{\partial_t \hat{\mathcal{S}}_t}$ for the isolated system from first principles, i.e. from the Liouville-von Neumann equation for the density matrix, using the standard generalized master equation approach of nonequilibrium statistical mechanics, \cite{vanHove,PrigogineR,Fujita,Nakajima,Zwanzig1} and by subsequently making reasonable assumptions regarding the factorization properties of the nonequilibrium probability distribution of microscopic states over the degrees of freedom of the different subsystems. 

We restrict here to weakly-coupled subsystems to show how our theory is consistent with the classical theory, to elucidate the manner in which the local equilibrium approximation can be fully abandoned, and to pave the way to study cases of strong coupling between subsystems for which the aforementioned factorization properties of the probability distribution of microscopic states become the main subject of study, marking a deep connection with quantum information theory. A detailed investigation of a new methodology to approach these cases will be considered elsewhere.

The pursue of the so outlined research program is essential both for a more fundamental understanding of nonequilibrium behavior,\cite{Jarzynski2} and because entropy production is inherent to dissipation so that a good atomic-scale description may have technological impact, e.g. by enabling  better control of waste heat and thermoelectric effects in single-molecule electronics,\cite{Aradhya,Lee,Pekola} guiding the efficient design of quantum refrigerators\cite{Feldmann} and quantum heat machines, \cite{Uzdin} nanosized photoelectric devices,\cite{Rutten} nanothermoelectric engines \cite{EspositoTM,EspositoL} based on quantum dots, etc., which are envisioned as practical applications of quantum thermodynamics. 

It turns out, as we show here with a particular example of electronic conduction in the presence of phonon modes playing the role of a reservoir, that our theory gives an explicit expression for the Joule heating from a calculation of the steady state electronic entropy production alone. This represents an important progress since this is done without calculating the rate of change of the energy of the electron subsystem.

This paper is organized as follows: in section \ref{phenterm} we give a brief review of entropy production and the second law of thermodynamics as they manifest in phenomenological thermodynamics. In section \ref{loceq} we discuss the local equilibrium assumption from a quantum perspective, with a derivation of the first law of thermodynamics from the expression for $\avg{\partial_t \hat{\mathcal{S}}_t}$ in this case, which is shown to hold for \emph{quasistatic} transformations or slow processes. This section, which mainly discusses how the foundations of the classical theory are to be understood quantum-mechanically, serves as a motivation to introduce the operator $\partial_t \hat{\mathcal{S}}_t$ for general isolated quantum systems of which possible reservoirs are part of. 

A transition is made in section \ref{masdS} to the generalized thermodynamic description of quantum systems. The second and third law of thermodynamics are established here for any isolated system, and an entropy balance equation is derived, splitting $\avg{\partial_t \hat{\mathcal{S}}}$ into entropy production and entropy transport terms. In section \ref{stochcl}, we show how the theory is consistent with Onsager's classical stochastic entropy production in an aged system. Finally, in section \ref{elPi} we calculate the electronic entropy production in a simple metal consisting of independent electrons weakly coupled to phonons in the presence of an external electric field,  deriving the Joule heating, and we conclude with section \ref{con}.

\section{Entropy production in phenomenological thermodynamics}\label{phenterm}
The thermodynamic definition of entropy \emph{changes} for any kind of process in a closed system (not interchanging particles with the reservoirs) was given by Clausius at the very end of his monumental 1865 paper. \cite{Clausius,Cropper} If the system, which is considered to be in contact with a set of heat sources at different temperatures $T$, follows a path $\gamma$ in the space of thermodynamic states, joining the initial and final arbitrary states $A$ and $B$, respectively, then the thermodynamic entropy change in the process is 
\begin{equation}\label{dSN}
\mathcal{S}_B-\mathcal{S}_A=N_C[\gamma]+ \int_{A}^{B}(\,\dbar Q/T)_{\gamma},
\end{equation}
where $\,\dbar Q$ is an infinitesimal amount of heat absorbed from (or surrendered to) the heat source at temperature $T$, and the quantity $N_C[\gamma]$, representing what came to be known as the ``uncompensated heat of Clausius'', \cite{Velasco} is a functional of the process. Clausius defined it in such a way that
\begin{equation}
N_C[\gamma]\equiv-\oint_\gamma \,\dbar Q/T=-\int_A^B(\,\dbar Q/T)_{\gamma}-\int_B^A(\,\dbar Q/T)_{\gamma_R}, 
\end{equation}
where $\gamma_R$ is an arbitrary \emph{reversible} path which is ``imagined'' to bring the system back to its initial state $A$. He proved that
\begin{equation}
N_C[\gamma]\geq0,\hspace*{0.5cm}\tm{(Clausius' inequality)}, 
\end{equation}
for any $\gamma$, which was a generalization of Carnot's results for cyclic processes; the equality holding if and only if $\gamma$ is a reversible path. This is the starting point of all the discussions found in textbooks of the second law of thermodynamics, \cite{Fermi} and is therefore regarded here as the fundamental expression for this law.

A classical formulation of nonequilibrium thermodynamics has been founded \cite{Prigogine,Groot} by taking as starting point \eqref{dSN} written in differential form and generalized to apply locally in small volume elements, $\delta v$, of a system 
\begin{equation}\label{dSei}
d\mathcal{S}=d_{\textrm{i}}\mathcal{S}+d_{\textrm{e}}\mathcal{S}, 
\end{equation}
where $d_{\textrm{i}}\mathcal{S}\equiv dN_C$ is the entropy produced, during an infinitesimal time interval, due to irreversible processes taking place inside the volume element, and $d_{\textrm{e}}\mathcal{S}$ the entropy supplied \emph{from} its surroundings ($\equiv \,\dbar Q/T$ for a closed element). The second law of thermodynamics requires only that the entropy produced satisfies
\begin{equation}\label{clau}
 d_{\textrm{i}}\mathcal{S}\geq0,  \hspace*{0.5cm}\tm{(Clausius' inequality)}.
\end{equation}
The theory so obtained for the phenomenological entropy production, $\Pi_{\delta v}=d_{\textrm{i}}\mathcal{S}/dt$, successfully describes \emph{slow} processes or phenomena where the decay time of local perturbations is very short compared to the global relaxation time, as in chemical reactions, diffusion processes, heat conduction, and their cross effects in gases and liquids. However, it requires fundamental modifications for \emph{fast} processes\cite{Jou2,Jou} and, in the following, we argue from a quantum-mechanical perspective why this happens to be the case, setting the stage and motivating the method for the subsequent development of our theory.

\section{Local equilibrium and quasistatic quantum transformations}\label{loceq}
Consider an isolated macroscopic system, possibly containing a set of particle and heat reservoirs which is divided into macroscopic subsystems. Microscopically, the total system is defined by the Hamiltonian
\begin{equation}\label{Ht}
\hat{H}=\sum_l \hat{H}_l+\sum_{l<m}\hat{H}_{lm},
\end{equation}
where $\hat{H}_l$ is the Hamiltonian of subsystem $l$, involving the kinetic energies of the particles comprising the subsystem as well as the energy of interaction among all these particles; and $\hat{H}_{lm}$ is the Hamiltonian respresenting the interactions among the particles of subsytem $l$ with those of subsystem $m$, possibly including hopping terms allowing particle transfer.

The fundamental assumption of statistical mechanics \cite{Landau} is that, since the interaction energy among the parts scales with their common surface areas, while the energy of the parts scales with their respective volumes, we can then remove all $\hat{H}_{lm}$ in \eqref{Ht} from a \emph{macroscopic} description of the dynamics and introduce instead a set of time-dependent parameters $\lbrace x_\lambda^l\rbrace$ embodying macroscopic constraints for the subsystem $l$, that evolve in time due to changes in the other subsystems. The operator representing macroscopic energy measurements in this approximation is
\begin{equation}\label{Hcal}
\hat{\mathcal{H}}=\sum_l\hat{\mathcal{H}}_l,\hspace{0.7cm}\tm{with}\hspace{0.3cm}\hat{\mathcal{H}}_l=\hat{H}_l(\lbrace x_\lambda^l\rbrace),
\end{equation}
where the notation in \eqref{Hcal} indicates that $\hat{\mathcal{H}}_l$ is to be taken as $\hat{H}_l$ plus an external potential due to the other subsystems and represented \emph{parametrically}.  For instance, a quantum subsystem acted upon by an external electric field is seen in the description of \eqref{Ht} as having a Coulomb potential energy (operator) coupling all the charges of the subsystem with all the charges outside of it which are sources of this field, while in the approximate description of \eqref{Hcal}, it is seen as coupled to an external parameter $\bm{E}$ representing the strength of the field. We shall call the latter the \emph{thermodynamic} description.

The local equilibrium assumption in the thermodynamic description is the statement that the macroscopic state of each part of our system, with a number of particles operator $\hat{\mathcal{N}}_l$, a temperature $T_l$ and a chemical potential $\mu_l$, is an equilibrium state. The local equilibrium density matrix of the total system is the factorized Gibbs state (subsystems macroscopically uncorrelated)
\begin{equation}\label{ler}
 \hat{\varrho}^{\textrm{r}}=\bigotimes_l\hat{\varrho}_l^{\textrm{r}}= \bigotimes_{l} \exp[-(\hat{\mathcal{H}}_l-\mu_l \hat{\mathcal{N}}_l-\Omega_l)/T_l],
\end{equation}
with  $\Omega_l=\Omega_l(T_l,\mu_l,\left\lbrace x_{\lambda}^l\right\rbrace)$ the thermodynamic potential of subsystem $l$, introduced so as to normalize the density matrix, that is, $\tm{Tr}\, \exp[-(\hat{\mathcal{H}}_l-\mu_l \hat{\mathcal{N}}_l)/T_l]=\exp(-\Omega_l/T_l)$. Note that, since the degrees of freedom of different subsystems are \emph{uncoupled} in the thermodynamic description, all operators $\hat{\mathcal{H}}_l$ and $\hat{\mathcal{N}}_m$ form a mutually commuting set and then define a natural basis of common eigenstates that we represent as $\lbrace\ket{\bm{\alpha}}\rbrace$.

The appearance of this natural set defines a family of observables acting on the system Hilbert space that, like $\hat{\mathcal{H}}_l$ and $\hat{\mathcal{N}}_m$, we call thermodynamic; these observables are diagonal in the basis $\lbrace\ket{\bm{\alpha}}\rbrace$. According to this, $\hat{\mathcal{H}}$ is a thermodynamic observable, and we denote the set of all these operators as 
\begin{equation}\label{fT}
\mathcal{T}=\lbrace\hat{\mathcal{G}}: [\hat{\mathcal{G}},\hat{\mathcal{H}}]=0\rbrace 
\end{equation}
Clearly all constant operators as well as all time-averaged observables\cite{Kubo,Zubarev} belong to this family. With $\mathcal{D}$ denoting the projection operator to the subspace spanned by $\lbrace\outpr{\bm{\alpha}}{\bm{\alpha}}\rbrace$, we can then split an arbitrary observable $\hat{G}$ in the convenient form
\begin{equation}
 \hat{G} = \mathcal{D}\hat{G}+\mathcal{N}\hat{G}=\hat{\mathcal{G}}+\hat{G}^{\sim},
\end{equation}
where $\hat{\mathcal{G}}=\mathcal{D}\hat{G}$ is the thermodynamically measurable part (or thermodynamic part) of $\hat{G}$; the complementary part being $\hat{G}^{\sim}=\mathcal{N}\hat{G}=\hat{G}-\hat{\mathcal{G}}$. 

The thermodynamic observables must have the characteristic of being slowly-varying quantities. \cite{Kirkwood,Green,Yamamoto,Zwanzig2} This is quantified in our theory by introducing a geometric measure, $\Delta$, of how approximate is the thermodynamic description. For this, let us introduce for an arbitrary observable $\hat{B}$ the hermitian operator
\begin{equation}
\hat{C}_{\mathcal{A}B}=-i[\hat{\mathcal{A}},\hat{B}] \hspace{0.5cm}\tm{with} \hspace{0.5cm}\hat{\mathcal{A}}\in \mathcal{T}, 
\end{equation}
and consider the simple geometry induced by the Hilbert-Schmidt norm $\lVert\hat{C}_{\mathcal{A}B}\rVert = (\textrm{Tr}\, \hat{C}_{\mathcal{A}B}^{\dagger}\hat{C}_{\mathcal{A}B})^{1/2}$. It is trivially seen that
\begin{equation}\label{c0}
\lVert\hat{C}_{\mathcal{H}0}\rVert=\lVert\hat{C}_{\mathcal{G}0}\rVert=0, 
\end{equation}
and by using the Jacobi identity for commutators, together with $[\hat{\mathcal{G}},\hat{\mathcal{H}}]=0$, coming from \eqref{fT}, we can write
\begin{equation}\label{jc}
\lVert\hat{C}_{\mathcal{H}\dot{\mathcal{G}}}\rVert=\lVert\hat{C}_{\mathcal{G}C_{\mathcal{H}H}}\rVert,
\end{equation}
where we identify $\dot{\hat{\mathcal{G}}}=\hat{C}_{\mathcal{G}H}$ when the parameters representing external constraints are fixed in time. Therefore, if $\hat{C}_{\mathcal{H}H}$ tend to the null operator in the norm, i.e. if we have $\Delta\rightarrow0$ with  
\begin{equation}
 \Delta^2 = \varepsilon_0^{-4}\lVert\hat{C}_{\mathcal{H}H}\rVert^2=\varepsilon_0^{-4}\sum_{\bm{\alpha}\bm{\alpha}'}(\varepsilon_{\bm{\alpha}}-\varepsilon_{\bm{\alpha}'})^2|\brket{\bm{\alpha}}{\hat{H}}{\bm{\alpha}'}|^2,
\end{equation}
where $\hat{\mathcal{H}}\ket{\bm{\alpha}}=\varepsilon_{\bm{\alpha}}\ket{\bm{\alpha}}$ and $\varepsilon_0$ is the smallest characteristic energy in the system (making $\Delta$ dimensionless); then we conclude, by using \eqref{c0}, \eqref{jc} as well as the continuity of the norm, that the quality of slow variation can be expressed as
\begin{equation}\label{gdot}
\lVert\dot{\hat{\mathcal{G}}}\rVert=\lVert [\hat{\mathcal{G}},\hat{H}]\rVert=O(\Delta). 
\end{equation}
The condition $\Delta\rightarrow0$ is physically realized when the thermodynamic limit is taken for all the subsystems comprising the total system, since in this limit $\avg{\hat{H}}$ and $\avg{\hat{\mathcal{H}}}$ tend to be indistinguishable for arbitrary states.

We now give the steps that constitute our general method in the next section. Given the density matrix $\hat{\rho}_t$ of the total system, we define the entropy operator as the negative of its logarithm, $\hat{S}_t=-\ln \hat{\rho}_t$ and, from this and the aforementioned discussion, the \emph{thermodynamic} entropy operator as $\hat{\mathcal{S}}_t=\mathcal{D}\hat{S}_t$. Since in the local equilibrium approximation the density matrix $\hat{\rho}_t=\hat{\rho}^{\tm{r}}$ is already diagonal in the basis $\lbrace\ket{\bm{\alpha}}\rbrace$, we have in this case
\begin{equation}\label{Sr}
 \hat{\mathcal{S}}^{\textrm{r}}=-\ln \hat{\varrho}^{\textrm{r}}=\dfrac{1}{T_l}\sum_l(\hat{\mathcal{H}}_l-\mu_l \hat{\mathcal{N}}_l-\Omega_l),
\end{equation}
where we have used the commutativity of all $\hat{\mathcal{H}}_l$ and $\hat{\mathcal{N}}_m$ to express $\ln\hat{\varrho}^{\textrm{r}}=\sum_l\ln\hat{\varrho}_l^{\textrm{r}}$. We are interested in thermodynamic entropy changes as the main observable, then the next step is an expression for $\avg{\partial_t \hat{\mathcal{S}}_t}$, which we get by first differentiating \eqref{Sr}
\begin{equation}\label{dSr}
\begin{split}
 d\hat{\mathcal{S}}^{\textrm{r}}= \sum_l\dfrac{1}{T_l}&\bigl[d\hat{\mathcal{H}}_l-d\mu_l \hat{\mathcal{N}}_l-\mu_l d\hat{\mathcal{N}}_l-d\Omega_l\\
&-\dfrac{dT_l}{T_l}(\hat{\mathcal{H}}_l-\mu_l \hat{\mathcal{N}}_l-\Omega_l)\bigr].
\end{split}
\end{equation}
Since $\Omega_l$ is a function of $\mu_l$, $T_l$ and of the external parameters, $x_{\lambda}^l$, implicit in $\hat{\mathcal{H}}_l$, we can differentiate the normalization relation $\tm{Tr}\, \exp[-(\hat{\mathcal{H}}_l-\mu_l \hat{\mathcal{N}}_l)/T_l]=\exp(-\Omega_l/T_l)$ after variations in these arguments to get, after noting that $\avg{\hat{G}_l}=\tm{Tr} \,\hat{\varrho}^{\textrm{r}} \hat{G}_l=\tm{Tr}\,\hat{\varrho}_l^{\textrm{r}} \hat{G}_l$ for a local operator $\hat{G}_l$ acting on the $l$-th subsystem,
\begin{equation}\label{dO}
 d\Omega_l= -\sum_{\lambda}F_{\lambda}^ldx_{\lambda}^l-\avg{\hat{\mathcal{N}}_l}d\mu_l-\dfrac{dT_l}{T_l}(\avg{\hat{\mathcal{H}}_l}-\mu_l\avg{\hat{\mathcal{N}}_l}-\Omega_l),
\end{equation}
with $F_{\lambda}^l=-\avg{\partial \hat{\mathcal{H}}_l/\partial x_{\lambda}^l}$ being the average force exerted by subsystem $l$ on its surroundings to get the displacements $dx_{\lambda}^l$. Taking expectation value of \eqref{dSr} and substituting \eqref{dO} we conclude that the average rate of change of the total thermodynamic entropy is in this case additive, $\avg{d\hat{\mathcal{S}}^{\textrm{r}}}=\sum_l \avg{d\hat{\mathcal{S}}_l^{\textrm{r}}}$, with
\begin{equation}\label{law1}
T_l\,\avg{d\hat{\mathcal{S}}_l^{\textrm{r}}}=\avg{d\hat{\mathcal{H}}_l}-\mu_l\avg{d\hat{\mathcal{N}}_l}+\sum_{\lambda}F_{\lambda}^ldx_{\lambda}^l,
\end{equation}
We have arrived in this way to the first law of thermodynamics, through a line of reasoning originally due to Gibbs, \cite{Gibbs} generalized here to the quantum case.

Note that for an arbitrary observable $\hat{G}$, the identity $\avg{d\hat{G}/dt}=\partial \avg{\hat{G}}/\partial t$ holds whenever the density matrix used to calculate the expectation value satisfies the Liouville-von Neumann equation, as is easily proved by changing to the Heisenberg picture within the expectation value operation, where $d\hat{G}/dt=\partial \hat{G}/\partial t-i[\hat{G},\hat{H}]$, with $\hat{G}$ depending explicitly on time in the Schr\"{o}dinger picture via the external parameters, and using the known identity $\textrm{Tr}\,\hat{A}[\hat{B},\hat{C}]=\textrm{Tr}\,\hat{C}[\hat{A},\hat{B}]$. Therefore, \emph{as long as} the local equilibrium density matrix $\hat{\rho}^{\tm{r}}$ satisfies the Liouville-von Neumann equation, we can commute the operation $\avg{d\hat{G}}=\partial\avg{\hat{G}}$ and write \eqref{law1} as the usual form of the first law of thermodynamics. 

The equivalence of \eqref{law1} with the usual form of the first law of thermodynamics 
\begin{equation}\label{tds}
T_l\,\partial_t\avg{\hat{\mathcal{S}}_l^{\textrm{r}}}=\partial_t\avg{\hat{\mathcal{H}}_l}-\mu_l\,\partial_t\avg{\hat{\mathcal{N}}_l}+\sum_{\lambda}F_{\lambda}^l\,\partial_t x_{\lambda}^l, 
\end{equation}
then requires that $\hat{\rho}^{\tm{r}}$ satisfies the Liouville-von Neumann equation $i\,\partial_t \hat{\varrho}^{\textrm{r}}= [\hat{H},\hat{\varrho}^{\textrm{r}}]$, where we use the symbol $\partial_t$ as a shorthand notation for $\partial/\partial t$. For this to be the case, it is necessary from \eqref{gdot} that 
\begin{equation}\label{rev}
\lVert \partial_t \hat{\rho}^{\tm{r}}\rVert=O(\Delta),
\end{equation}
since $ \hat{\rho}^{\tm{r}}$ is expressed in terms of thermodynamic observables. When the thermodynamic limit is taken for each subsystem, we have $\Delta\rightarrow0$, and then the parameters $x_{\lambda}^l$ should vary with time so slowly that the state $\hat{\varrho}^{\textrm{r}}$ can be interpreted as ``moving'' in a locus of equilibrium states, so that $\lVert\partial_t\hat{\varrho}^{\textrm{r}}\rVert\rightarrow0$ in \eqref{rev}. These are precisely the \emph{quasistatic} (or reversible) transformations for which the first law involving thermodynamic entropy changes applies, hence the superindex ``r'' standing for reversible, and the systematic omission of the time subindex in the variables. Note that in this case, the quantity $N_C[\gamma]$ in \eqref{dSN} vanishes for any $\gamma=\lbrace x_\lambda^l(t), \forall\, \lambda,l\;\tm{and}\; t\in[t_A,t_B]\rbrace$.

A nonzero entropy production appears instead when the subsystems are macroscopic at the atomic scale, but compared to the size of the total system, they are small volume elements, $\delta v_l$. In this case, an entropy balance equation may be obtained from \eqref{tds}, by using the relations
\begin{eqnarray}
  \partial_t\avg{\hat{\mathcal{H}}_l}&=&-\sum_m J_{\mathcal{H}}^{lm}+\sum_{\lambda}\dfrac{\partial\avg{\hat{\mathcal{H}}_l}}{\partial x_{\lambda}^l}\partial_t x_{\lambda}^l,\label{c1}\\
  \partial_t\avg{\hat{\mathcal{N}}_l}&=&-\sum_m J_{\mathcal{N}}^{lm}+\sum_{\lambda}\dfrac{\partial\avg{\hat{\mathcal{N}}_l}}{\partial x_{\lambda}^l}\partial_t x_{\lambda}^l,\label{c2}
\end{eqnarray}
which state that the average macroscopic energy and number of particles of a given subsystem can only change by transport to other subsystems, defining the corresponding currents $J_{\mathcal{H}}^{lm}$ and $J_{\mathcal{N}}^{lm}$ in terms of quantities proportional to the particle velocities, with an appropriate microscopic account for the heat currents, plus terms allowing the technical possiblity of a creation or destruction of particles induced by the variation of the external constrainsts. Substituting these in \eqref{tds} we get
\begin{equation}\label{cPiPhi}
\begin{split}
 \partial_t\avg{\hat{\mathcal{S}}_l^{\textrm{r}}}&=\dfrac{1}{T_l}\sum_{\lambda}\left[\,\dfrac{\partial}{\partial x_{\lambda}^l}(\avg{\hat{\mathcal{H}}_l}-\mu_l\avg{\hat{\mathcal{N}}_l})+F_{\lambda}^l\,\right]\partial_t x_{\lambda}^l\\
 \ &-\dfrac{1}{T_l}\sum_m\left(J_{\mathcal{H}}^{lm}-\mu_l J_{\mathcal{N}}^{lm}\right)= \Pi_{\delta v_{l}}-\Phi_{\delta v_{l}},
 \end{split}
\end{equation}
the first term in the first equality being the entropy production term, $\Pi_{\delta v_{l}}$, and the second one the entropy transport term, $\Phi_{\delta v_{l}}$. Results consistent with the classical theory are obtained when particle creation or destruction is not observed macroscopically, in which case \eqref{c1} and \eqref{c2} are just the usual conservation laws (continuity equations) and the entropy production in the subsystem reduces to the well-known sum of products of thermodynamic forces times the rate of change of their conjugate external parameters
\begin{equation}\label{Picl}
\Pi_{\delta v_{l}} = \dfrac{1}{T_l}\sum_{\lambda} F_{\lambda}^l\,\partial_t x_{\lambda}^l,\hspace{0.5cm}\tm{(classical).}
\end{equation}
The presentation given here can be straightforwardly generalized by considering local equilibrium Gibbs ensembles more general than \eqref{ler}, that is, by augmenting the thermodynamic entropy operator \eqref{Sr} with terms proportional to the components of the macroscopic linear and angular momentum operators of each subsystem, \cite{Landau} with \eqref{c1} and \eqref{c2} expanded to include the conservation laws of their respective expectation values.

Note that we have kept the superindex ``r'' (although not strictly with its original connotation) in \eqref{cPiPhi} because, even though the thermodynamic limit is not taken for each susbsytem, which would make $\Delta\rightarrow0$ and the processes necessarily quasistatic, the fact that the volume elements, $\delta v_l$, are macroscopic at the atomic scale still implies that $\Delta$ is very small and hence, from \eqref{rev}, that the variations $\partial_t \hat{\rho}^{\tm{r}}$ should correspondingly be very small in the norm. As mentioned in section \ref{phenterm}, we then see why the classical theory works well for \emph{slow} processes, i.e. those for which the time to get relaxation to equilibrium within each volume element is much shorter than the time to get equilibrium among them. 

The discussion in this section elucidates the problems with the local equilibrium assumption and previous theories of entropy production, which rely on expressions of the type \eqref{tds} together with conservation laws, like \eqref{c1} and \eqref{c2}, as in the classical theory. As we have made explicit, developing a theory of entropy production from \eqref{tds} inherently assumes that the correlations among the subsystems of a large isolated system are negligible \emph{for all times}, and using \eqref{c1} in this theory takes for granted that an appropriate mechanical description of the microscopics of heat currents have been univocally achieved. 

We now propose a way to derive an entropy balance equation for the subsystems of a general isolated system from first principles, starting from the Liouville-von Neumann equation for the density matrix of the isolated system, which does not rely on the above assumptions. 

\section{Master equation for the thermodynamic entropy operator}\label{masdS}
We generalize the thermodynamic description to include subsystems which are not distinguished by spatial boundaries and which are not necessarily macroscopic at the atomic scale. The key point to borrow from thermodynamics is the existence of the \emph{thermodynamic basis} $\lbrace\ket{\bm{\alpha}}\rbrace$ and the interpretation of thermodynamic observables as those which are diagonal in this basis.  That is, we consider an isolated quantum system (containing possible reservoirs) which has a Hamiltonian $\hat{\mathcal{H}}$ representing the energy of uncoupled subsystems, as before, and study the dynamics when the perturbation, $\hat{V}$, mixing the degrees of freedom of the different subsystems, or a set of them, is turned on. 

The Hamiltonian of the total system is then given by $\hat{H}=\hat{\mathcal{H}}+\hat{V}$, and the situations of interest include phenomena such as quantum quenches, \cite{Polkovnikov,Yukalov,DAlessio,GogolinE,EisertFG} or the response to applied fields. \cite{Kubo,Suzuki,Suzuki2} After preparation of the system in an initial statistical state of the form
\begin{equation}
\hat{\rho}_0=\exp(-\hat{S}_0), 
\end{equation}
with $\hat{S}_0$ an arbitrary (in general unbounded) hermitian operator with $[\hat{S}_0,\hat{V}]\neq0$, the nonequilibrium state is described by the evolved density matrix $\hat{\rho}_t$, and we define the entropy operator $\hat{S}_t$ by 
\begin{equation}
\hat{\rho}_t=\exp(-\hat{S}_t),\hspace{0.5cm}\textrm{or}\hspace{0.5cm}\hat{S}_t=-\ln \hat{\rho}_t,
\end{equation}
which can always be written since the density matrix is positive-definite. This exponential representation of the density matrix is not new; it is a generalized form \cite{Suzuki1,Suzuki2} of the nonequilibrium statistical operator introduced by Zubarev, \cite{Zubarev,Zubarev2} and obtained for the case of steady states by Hershfield \cite{Hershfield}.

As discussed in the previous section, our \emph{new} thermodynamic entropy operator, $\hat{\mathcal{S}}_t=-\mathcal{D}\ln \hat{\rho}_t$, is obtained from $\hat{S}_t$ by projecting to the space of operators diagonal in the basis $\lbrace\ket{\bm{\alpha}}\rbrace$ of eigenstates of $\hat{\mathcal{H}}$. We now establish the second law of thermodynamics for nonequilibrium transformations of the total system. For this, we consider for simplicity the specific situation of initial states diagonal in the thermodynamic basis, e.g. those of local equilibrium form as in \eqref{ler}, for which $\hat{S}_0^{\sim}=0$ or $\hat{S}_0=\hat{\mathcal{S}}_0$. These initial states are usually assumed in practice \cite{Mori2,Kubo,Hershfield}, e.g. in transport problems.

Let us denote the diagonal (or thermodynamic) part of the density matrix of the system as
\begin{equation}
\hat{\varrho}_t=\mathcal{D}\hat{\rho}_t. 
\end{equation}
The occupation probability of the state $\ket{\bm{\alpha}}$ is obtained by taking matrix elements $P_{\bm{\alpha};t}=\brket{\bm{\alpha}}{\hat{\varrho}_t}{\bm{\alpha}}$. The proof now follows in steps by first using a corollary to Klein's inequality, \cite{Wehrl} which states that for any \emph{concave} function $f(x)$ we have
\begin{equation}
 \textrm{Tr}\;f(\hat{\varrho}_t)\geq \textrm{Tr}\;f(\hat{\rho}_t).
\end{equation}
By choosing the concave function $f(x)=-x\ln(x)$, we easily get
\begin{equation}\label{ie1}
 -\textrm{Tr}\;\hat{\varrho}_t\ln(\hat{\varrho}_t)\geq -\textrm{Tr}\;\hat{\rho}_t\ln\hat{\rho}_t\hspace{0.5cm}\textrm{or}\hspace{0.5cm}S_{d;t}\geq S_{vN;t},
\end{equation}
where we have denoted $S_{d;t}=-\sum_{\bm{\alpha}}P_{\bm{\alpha};t}\ln P_{\bm{\alpha};t}$ as the diagonal entropy \cite{Polkovnikov, Polkovnikov2,Santos,Levi} and $S_{vN;t}$ is the well-known von Neumann entropy. Using the time-invariance of $S_{vN;t}$ under the unitary evolution of the isolated system together with the fact that the initial state is diagonal, so that $S_{d;0}=S_{vN;0}$, then \eqref{ie1} implies \cite{Polkovnikov,Polkovnikov2}
\begin{equation}
S_{d;t}\geq S_{d;0}.
\end{equation}
We use this result and the Husimi-Mori lemma, \cite{Husimi,Mori1} which states that for any \emph{convex} function $g(x)$ and state $\ket{\psi}$ we have
\begin{equation}
\brket{\psi}{g(\hat{\rho}_t)}{\psi} \geq g(\brket{\psi}{\hat{\rho}_t}{\psi}),
\end{equation}
to show that, if we choose the convex function $g(x)=-\ln(x)$ so that $-\brket{\bm{\alpha}}{\ln\hat{\rho}_t}{\bm{\alpha}}\geq -\ln P_{\bm{\alpha};t}$, the thermodynamic entropy satisfies
\begin{equation}\label{cst}
\mathcal{S}_t=\avg{\hat{\mathcal{S}}_t}= -\sum_{\bm{\alpha}}P_{\bm{\alpha};t}\brket{\bm{\alpha}}{\ln\hat{\rho}_t}{\bm{\alpha}}\geq S_{d;t}\geq \mathcal{S}_0,
\end{equation}
where $\mathcal{S}_0=S_{d;0}=S_{vN;0}$ by the assumption of the initial diagonal state. For our isolated system for which there is no entropy to be transported outside of its boundaries, this proves that $\mathcal{S}_t$ satisfies the second law of thermodynamics. 

Note that, by splitting $\hat{\rho}_t=\hat{\varrho}_t+\hat{\rho}_t^{\sim}$ and using the convenient resolvent representation of the logarithm of an operator sum \cite{Suzuki1}
\begin{equation}
 \ln(\hat{A}+\hat{B})=\int_0^{\infty}dx\left(\dfrac{1}{x+1}-\dfrac{1}{x +\hat{A}+\hat{B}}\right),
\end{equation}
we can expand the thermodynamic entropy as 
\begin{equation}\label{stsd}
\begin{split}
 \mathcal{S}_t&=\;S_{d;t}\\
 &+\;\sum_{\bm{\alpha},\,\bm{\beta}(\neq \bm{\alpha})}\left[\dfrac{1}{(P_{\bm{\beta};t}-P_{\bm{\alpha};t})}-\dfrac{P_{\bm{\alpha};t}}{(P_{\bm{\beta};t}-P_{\bm{\alpha};t})^2}\ln\dfrac{P_{\bm{\beta};t}}{P_{\bm{\alpha};t}}\right]\\
 &\ \hspace{2cm}\times|\brket{\bm{\alpha}}{\,\hat{\rho}^{\sim}\,}{\bm{\beta}}|^2\;+\;O(\brket{}{\hat{\rho}^{\sim}}{}^3),
\end{split}
\end{equation}
with $ \mathcal{S}_t-S_{d;t}\geq 0$ due to \eqref{cst}; therefore the thermodynamic entropy, unlike the diagonal entropy, is able to capture entropy increasing processes due to quantum correlations or entanglement among the different subsystems, that are encapsulated in the off-diagonal elements of the density matrix. When these quantum correlations are negligible which, as discussed in section \ref{loceq}, is the case when each subsystem is macroscopic, the diagonal entropy \emph{becomes} the thermodynamic entropy according to \eqref{stsd}, and due to the quasistatic (or slow) nature of the global transformations involved in this case, the thermodynamic basis may be referred to as the adiabatic basis. \cite{Polkovnikov_ab,Polkovnikov} 

The thermodynamic entropy, unlike the diagonal and von-Neumann entropies, satisfies the third law of thermodynamics in a transparent way. The third law states that the thermodynamic entropy at zero temperature must be zero. The standard argument is that at zero temperature any physical state is \emph{pure}. For an arbitrary pure state $\ket{\psi}$, there is always an orthonormal basis of Hilbert space which has this state as one of its elements (construct it via the Gram-Schmidt procedure starting from $\ket{\psi}$). Denote this basis, $\lbrace\ket{\psi_r}\rbrace$, and order its elements such that $\ket{\psi}=\ket{\psi_1}$. We take this basis as the reference for ``diagonal''. With this we then have for the diagonal and von-Neumann entropies
\begin{equation}\label{Sdp}
\begin{split}
S_d(\psi)&=S_{vN}(\psi)=-\sum_r P_r \ln P_r, \\
&= -1\cdot\ln(1)-\sum_{r\neq 1}0\cdot\ln(0),
\end{split}
\end{equation}
where $P_r$ is the probability that the system be found in state $\ket{\psi_r}$. Eq. \eqref{Sdp} is usually \emph{understood} to be zero, \cite{Wehrl} although it is clearly an undetermined quantity since, taken at face value, $-0\cdot\ln(0)=0\cdot\infty$. 

The thermodynamic entropy of pure states is well-defined and readily vanishes. In order to show this, we denote the density matrices (projectors) $\hat{\rho}_r=\outpr{\psi_r}{\psi_r}$, with $\sum_r\hat{\rho}_r=\hat{1}$. We can then write
\begin{equation}\label{l1}
 \ln \hat{\rho}_1 = \ln(\hat{1}-\textstyle{\sum_{r\neq 1}}\hat{\rho}_r)=-\sum_{u=1}^{\infty}(\textstyle{\sum_{r\neq 1}}\hat{\rho}_r)^u/u.
\end{equation}
Using this, we can compute the thermodynamic entropy of the state $\ket{\psi}$ as
\begin{equation}
 \mathcal{S}(\psi)= -\sum_r \brket{\psi_r}{\hat{\rho}_1\mathcal{D}\ln(\hat{\rho}_1)}{\psi_r}=-\brket{\psi}{\ln(\hat{\rho}_1)}{\psi}.
\end{equation}
This clearly vanishes exactly since $\ket{\psi}=\ket{\psi_1}$ is orthogonal to all $\ket{\psi_{r\neq1}}$ involved in the last equality of \eqref{l1}. This establishes the third law of thermodynamics.

We are after an entropy balance equation for the subsystems, so we need an equation of motion for $\hat{\mathcal{S}}_t$ and a procedure to get from this one for each subsystems, as in the previous section. This can be obtained by first noting that the usual unitary evolution of the density matrix implies that $\hat{S}_t$ also satisfies the Liouville-von Neumann equation \cite{Suzuki1} satisfied by $\hat{\rho}_t$. We have
\begin{equation}\label{lS}
 i\partial_t \hat{S}_t=[\hat{H},\hat{S}_t]\equiv L\,\hat{S}_t.
\end{equation}
This allows us to follow exactly the same procedure originally used with the density matrix \cite{Nakajima,Zwanzig1} to derive an equation of motion for its diagonal part, $\hat{\varrho}_t$, the so-called Nakajima-Zwanzig generalized master equation. That is, we split the entropy operator into a diagonal and nondiagonal part, with respect to the eigenbasis of $\hat{\mathcal{H}}$, as $\hat{S}_t=\hat{\mathcal{S}}_t+\hat{S}_t^{\sim}$, and obtain an equation of motion for the diagonal part using Zwanzig's integral \cite{Zwanzig1}
\begin{equation}\label{dSdt}
 i \partial_t\hat{\mathcal{S}}_t=\mathcal{D}L\hat{\mathcal{S}}_t+\mathcal{D}L\,e^{-it\mathcal{N}L}\hat{S}_0^{\sim}
 -i\int_0^td\tau K_{\tau}\,\hat{\mathcal{S}}_{t-\tau},
\end{equation}
where the memory kernel is defined as \cite{Zwanzig3}
\begin{equation}\label{mk}
 K_{\tau}=\mathcal{D}L\,e^{-i\tau\mathcal{N}L}\mathcal{N}L.
\end{equation}
Now, it is easy to verify that $\mathcal{D}L \mathcal{D}=0$ for any Hamiltonian, \cite{Zwanzig3} therefore the first term in \eqref{dSdt} vanishes and, with our initial diagonal states implying $\hat{S}_0^{\sim}=0$, we are left with the integro-differential equation
\begin{equation}\label{conv}
 \partial_t\hat{\mathcal{S}}_t=-\int_0^td\tau K_{\tau}\,\hat{\mathcal{S}}_{t-\tau}.
\end{equation}
Although an exact solution for \eqref{conv}, as well as for the similar equation satisfied by $\hat{\varrho}_t$, can easily be found by a Laplace transformation followed by an inversion
\begin{equation}\label{St}
\hat{\mathcal{S}}_t =\dfrac{1}{2\pi i}\int_{c-i\infty}^{c+i\infty} ds \dfrac{e^{st}}{s+K_s}\hat{\mathcal{S}}_0, \hspace{0.5cm}\textrm{
with}\;\; c>0,
\end{equation}
where $K_s$ is the Laplace transform of the memory kernel, obtained from \eqref{mk} as
\begin{equation}\label{ker}
 K_s=\mathcal{D}L\,\dfrac{1}{s+i\mathcal{N}L}\mathcal{N}L,
\end{equation}
we restrict here, for the sake of a clear presentation and for comparison with the classical results, to the Born-Markov approximation for \emph{weakly} coupled subsystems, leaving a more general discussion for another publication. This approximation, which is justified in the limit of very weak coupling potentials, $\hat{V}$, and very long times (Van Hove limit \cite{vanHove2,Davies}) amounts to neglecting memory effects in \eqref{conv}. In practice, this works for times \emph{after} any transient effect or prethermalization plateau \cite{Stark,Bertini,Nessi} of the isolated system has passed. We then have in this limit
\begin{equation}\label{dsM}
 \partial_t\hat{\mathcal{S}}_t=-\lim_{s\rightarrow0^{+}}K_{s}\,\hat{\mathcal{S}}_{t},
\end{equation}
where $K_s$ and $\hat{\mathcal{S}}_{t}$, after being expanded in powers of $\hat{V}$, are truncated up to the lowest orders, for which the well-known identity for the resolvent operator expansion
\begin{equation}
(A+B)^{-1}=A^{-1}-A^{-1}B\,(A+B)^{-1}, 
\end{equation}
is very useful. Taking expectation value of \eqref{dsM}, and noting that for a diagonal operator $\hat{\mathcal{G}}$ we have $\avg{\hat{\mathcal{G}}}_t= \textrm{Tr}\, \hat{\rho}_t\hat{\mathcal{G}}=\textrm{Tr}\, \hat{\varrho}_t\hat{\mathcal{G}}$, the average rate of change of the thermodynamic entropy in the Born-Markov limit is then
\begin{equation}\label{bdS}
\avg{\partial_t\hat{\mathcal{S}}_t}=\sum_{\bm{\alpha}\bm{\alpha}'}P_{\bm{\alpha}}W_{\bm{\alpha}\bm{\alpha}'}\ln\frac{P_{\bm{\alpha}}}{P_{\bm{\alpha}'}}.
\end{equation}
with the transition rates $W_{\bm{\alpha}\bm{\alpha}'}=2\pi\delta(\varepsilon_{\bm{\alpha}}-\varepsilon_{\bm{\alpha}'})|V_{\bm{\alpha}\bm{\alpha}'}|^2$, calculated in the lowest order in the coupling potential using Fermi's golden rule. Here, we have derived the transition rates from \eqref{ker} and \eqref{dsM}, by using the representation of the delta function \cite{Lippmann}
\begin{equation}
\lim_{s\rightarrow0^{+}}\textrm{Re}\,\dfrac{1}{s+i\omega}=\pi\delta(\omega). 
\end{equation}
Moreover, $P_{\bm{\alpha}}=\brket{\bm{\alpha}}{\hat{\varrho}_t^{(0)}}{\bm{\alpha}}$ is the occupation probability of the state $\ket{\bm{\alpha}}$ in its lowest-order approximation, \cite{Nakajima} which also satisfies the Born-Markov limit of the generalized master equation, that is, the transport (or Pauli) equation
\begin{equation}\label{pau}
 \partial_t P_{\bm{\alpha}}= \sum_{\bm{\alpha}'}(P_{\bm{\alpha}'}W_{\bm{\alpha}'\bm{\alpha}}-P_{\bm{\alpha}}W_{\bm{\alpha}\bm{\alpha}'}). 
\end{equation}
The right hand side of \eqref{bdS} can be rearranged to yield the quantum version, in the Born-Markov limit, of the entropy balance equation. We find
\begin{equation}\label{balS}
 \avg{\partial_t\hat{\mathcal{S}}_t}=\Pi-\Phi,
\end{equation}
where the average rate of entropy produced in the system is interpreted as \cite{Schnakenberg,Luo,Seifert,EspositoB,Tome}
\begin{equation}\label{Pi}
 \Pi=\dfrac{1}{2}\sum_{\bm{\alpha},\bm{\alpha}'}\left(P_{\bm{\alpha}}W_{\bm{\alpha}\bm{\alpha}'}-P_{\bm{\alpha}'}W_{\bm{\alpha}'\bm{\alpha}}\right)\ln\dfrac{P_{\bm{\alpha}}W_{\bm{\alpha}\bm{\alpha}'}}{P_{\bm{\alpha}'}W_{\bm{\alpha}'\bm{\alpha}}},
\end{equation}
and the average entropy flux to the surroundings as
\begin{equation}\label{Phi}
\Phi=\dfrac{1}{2}\sum_{\bm{\alpha},\bm{\alpha}'}\left(P_{\bm{\alpha}}W_{\bm{\alpha}\bm{\alpha}'}-P_{\bm{\alpha}'}W_{\bm{\alpha}'\bm{\alpha}}\right)\ln\dfrac{W_{\bm{\alpha}\bm{\alpha}'}}{W_{\bm{\alpha}'\bm{\alpha}}}.
\end{equation}
Of course, the latter must be zero for an isolated system since a global entropy current finds nowhere to go in this case. The vanishing of this quantity is clearly seen from the symmetry of the transition rates $W_{\bm{\alpha\alpha'}}$ under the interchange of indices, resulting from the hermiticity of the perturbation $\hat{V}$. A nonvanishing entropy current is obtained however when we consider the local entropy production in a subsystem of a larger system, as in the electrical conduction problem of section \ref{elPi}.

Note that $\Pi$ is a sum of terms of the form $(x-y)\ln(x/y)$ so  is always \emph{non-negative}. It vanishes for reversible transformations (local equilibrium) or in equilibrium due to detailed balance, $P_{\bm{\alpha}}^{\textrm{r}}\,W_{\bm{\alpha}\bm{\alpha}'}=P_{\bm{\alpha}'}^{\textrm{r}}\,W_{\bm{\alpha}'\bm{\alpha}}$, this being a statistical statement of the second law of thermodynamics in the Clausius form. The outlined method is the one that we shall follow in section \ref{elPi} for the electrical conduction problem to derive an entropy balance equation for the electronic subsystem in the Born-Markov limit, based on the transport equation for the total electrons $+$ phonons $+$ field system, without any need to invoke expressions like \eqref{tds} together with extra conservation laws.

One of the advantages of our approach, besides being grounded on fundamental facts regarding the nature of thermodynamic observables is that, as opposed to actively studied relative-entropy formulations \cite{Spohn2,Spohn,EspositoNJP} of quantum entropy production, it can be generalized to initial states with correlations among the subsystems, i.e. not of the local equilibrium form. This is very important since the neglection of correlations in the state of an isolated system is inconsistent with the specification of its energy. \cite{Philippot} We have safely ignored this fact in our present discussion because the consideration of a nonvanishing second term in \eqref{dSdt}, due to $\hat{S}_0^{\sim}\neq0$, only adds the term
\begin{equation}
 \dfrac{1}{2\pi i}\int_{c-i\infty}^{c+i\infty} ds\dfrac{e^{st}}{s+K_s}\mathcal{D}L\dfrac{1}{s+i\mathcal{N}L}\hat{S}_0^{\sim},
\end{equation}
to the solution \eqref{St}. However, it is easily seen that expressions containing $\hat{S}_0^{\sim}$ contribute higher order terms in the weak coupling expansion embodied in the Born-Markov limit and then are negligible; the same happens \cite{Nakajima} for the contributions coming from $\hat{\rho}_0^{\sim}$ in the Born-Markov limit of the generalized master equation for $\hat{\varrho}_t$. Therefore, our formalism has room to study memory effects and strong correlations in the initial state by only straightforward modifications. These memory effects are the ones responsible for heat transport depending on the path of thermodynamic states in phenomenological thermodynamics.

\section{Relation with classical stochastic thermodynamics}\label{stochcl}
We now show that our result, \eqref{bdS}, is consistent with the result for the average rate of change of the thermodynamic entropy obtained in Onsager's classical theory. We consider an isolated macroscopic system which has been left alone for a very long time (aged system). The classical thermodynamic state is described by a set of extensive variables, such as energy, mass, electric charge, etc., which randomly fluctuate about their equilibrium values and whose values define the  classical state of the system. This state is represented by the symbol $\bm{a}_t$ (shifted to vanish in equilibrium), whose successive values in time describe a stationary stochastic process. 

It can be shown that, if the fluctuations follow a Gaussian process, which can be argued to be the case if the extensive variables are algebraic sums of very many independent (weakly coupled) ``microscopic'' quantities so that the central limit theorem can be invoked and, if in addition the process is Markovian, then the joint probability distribution,
\begin{equation}\label{jD}
 \Omega(\bm{a}',\Delta t,\bm{a}'')=P_{\bm{a}'}P_{\bm{a}'\bm{a}''}(\Delta t),
\end{equation}
for observing the values $\bm{a}_{t'}=\bm{a}'$ and $\bm{a}_{t''}=\bm{a}''$ at the respective times separated by an interval $\Delta t=t''-t'$, with $P_{\bm{a}'\bm{a}''}(\Delta t)$ the corresponding conditional probability to make a transition between these states,  is given by Onsager's principle, \cite{Onsager2,Hashitsume,Onsager3} which we write as \footnote{We are interpreting the normal forward evolution as the mirror image in time of that originally considered by Onsager and Machlup since, as pointed out by them, the mirror image is the only one that satisfies the initial conditions nontrivially.} 
\begin{equation}\label{Ons}
  2\ln \Omega(\bm{a}',\Delta t,\bm{a}'')= \mathcal{S}_{\bm{a}'}+\mathcal{S}_{\bm{a}''}+\left(\int_{t'}^{t''}d\tau\,\dot{\mathcal{S}}\right)_{\textrm{min}},
\end{equation}
where the path of integration is the trajectory, $\bm{a}_{\tau}$, which makes the integral a minimum, subject to the conditions $\bm{a}_{t'}=\bm{a}'$ and $\bm{a}_{t''}=\bm{a}''$. Clearly, if we take the limit $\Delta t\rightarrow 0$, the integral tends to $\dot{\mathcal{S}}_{\bm{a}'}\Delta t$, where $\dot{\mathcal{S}}_{\bm{a}'}$ is the entropy production rate in the state  $\bm{a}'$, whose entropy is related to the probability distribution, $P_{\bm{a}'}$, by Boltzmann's principle. Subtracting the time-reversed expression of Onsager's principle from \eqref{Ons} we get, in the limit $\Delta t\rightarrow0$, the alternative form
\begin{equation}\label{lf}
 \ln \dfrac{\Omega(\bm{a}',\Delta t,\bm{a}'')}{\Omega(\bm{a}'',-\Delta t,\bm{a}')}=\dfrac{1}{2}(\dot{\mathcal{S}}_{\bm{a}'}+\dot{\mathcal{S}}_{\bm{a}''})\Delta t,
\end{equation}
We now average \eqref{lf} over the joint distribution \eqref{jD}, which is expanded up to linear order in $\Delta t$ by writing the transition probabilities to go from $\bm{a}'$ to $\bm{a}''$ after a time $\Delta t$ as
\begin{equation}
 P_{\bm{a}'\bm{a}''}(\Delta t)=\delta_{\bm{a}'\bm{a}''}+W_{\bm{a}'\bm{a}''}\Delta t=P_{\bm{a}''\bm{a}'}(-\Delta t),
\end{equation}
the last equality being the statement of Onsager's microscopic reversibility, \cite{Onsager2,Casimir} and leading to the symmetry of the transition rates, $W_{\bm{a}'\bm{a}''}$, under the interchange of indices. This symmetry allows to write the averaged left-hand side of \eqref{lf} as $\Delta t\, \sum_{\bm{a}'\bm{a}''}P_{\bm{a}'}W_{\bm{a}'\bm{a}''}\ln(P_{\bm{a}'}/P_{\bm{a}''})$ and the right-hand side as $\Delta t\,\sum_{\bm{a}'}P_{\bm{a}'}\dot{\mathcal{S}}_{\bm{a}'}$ Therefore, by recognizing the latter sum as $\avg{\dot{\mathcal{S}}}$, we get the expression
\begin{equation}\label{cdS}
  \avg{\mathcal{\dot{S}}}=\sum_{\bm{a}'\bm{a}''}P_{\bm{a}'}W_{\bm{a}'\bm{a}''}\ln \dfrac{P_{\bm{a}'}}{P_{\bm{a}''}},
\end{equation}
which gives the desired link with our theory, by comparing with \eqref{bdS}. We remark that \eqref{lf} is of the same form \cite{Bustamante} 
\begin{equation}
 \ln\dfrac{P_{\Delta t}(\sigma)}{P_{\Delta t}(-\sigma)}=\sigma\Delta t
\end{equation}
as Gallavotti and Cohen fluctuation theorem, \cite{GC} if we read $(1/2)(\dot{\mathcal{S}}_{\bm{a}'}+\dot{\mathcal{S}}_{\bm{a}''})$, as a realization of the random number $\sigma = (1/2)(\dot{\mathcal{S}}_{\bm{a}_{t'}}+\dot{\mathcal{S}}_{\bm{a}_{t''}})$, representing the average entropy production in going from $\bm{a}_{t'}$ to $\bm{a}_{t''}$ during a time interval $\Delta t$ along the stochastic trajectory of states; and translate the joint probability, $\Omega(\bm{a}',\pm\Delta t,\bm{a}'')$, to have the state realizations $\bm{a}_{t'}=\bm{a}'$ and $\bm{a}_{t''}=\bm{a}''$, in a forward ($+\Delta t$) or backward ($-\Delta t$) evolution, to the corresponding probabilities $P_{\Delta t}(\pm\sigma)$ to have the realization, $(1/2)(\dot{\mathcal{S}}_{\bm{a}'}+\dot{\mathcal{S}}_{\bm{a}''})$, of $\sigma$ or its time-reversed value.

\section{Entropy production in electrical conduction}\label{elPi}
We next apply the formalism to a model of independent electrons coupled to phonons in the presence of an electric field. We are interested in the average rate of entropy produced in the electronic system and transported to the phonons in the steady state. The picture is then that of a large system divided into three subsystems, the electrons, the phonons, and the sources of the field. In the thermodynamic description we parametrize, as usual, the coupling to the latter by introducing $\bm{E}_t$ and forgetting about the structure of this subsystem.

The Hamiltonian of the total system is then
\begin{equation}
 \hat{H}=\hat{H}_{\tm{el}}+\hat{H}_{\tm{ph}}+\hat{H}_{\tm{el-ph}}+\hat{H}_{F;t},
\end{equation}
where $\hat{H}_{\tm{el}}=\sum_k\epsilon_k\,\hat{c}_k^{\dagger}\hat{c}_k$ is the kinetic energy operator for the electrons, which are assumed to be  free except for their interaction with the field and the phonons, the energy operator of the phonon subsystem is $\hat{H}_{\tm{ph}}=\sum_q\omega_q\,\hat{a}_q^{\dagger}\hat{a}_q$,  and the electron-phonon interaction is bilinear in electron operators and linear in phonon operators
\begin{equation}\label{Helph}
 \hat{H}_{\tm{el-ph}}=\sum_{qkk^\prime}M_{k'k}^q\,\hat{c}_{k'}^{\dagger}\hat{c}_{k}\,(\hat{a}_{q}+\hat{a}_{-q}^{\dagger}),
\end{equation}
with $M_{k'k}^q$ representing the strength of the coupling. The generalization to multiple electronic bands and multiple phonon branches is straightforward and does not change the results. Finally, $\hat{H}_{F;t}$ represents the effects of the applied electric field, $\bm{E}_t$, and can be written in first-quantized notation as $\hat{H}_{F;t}=-e\,\bm{E}_t\cdot \sum_e\hat{\bm{x}}_e$, where  $\hat{\bm{x}}_e$ is the  displacement of electron $e$ from some arbitrarily chosen reference position.

Up to time $t=0$ we have a collection of electrons in local equilibrium with the lattice vibrations of a metal at a temperature $T$, and no applied electric electric field, i.e. $\bm{E}_0=0$. The initial state is then of the form
\begin{equation}\label{ep0}
\hat{\rho}_0=Z^{-1}\exp[-(\hat{\mathcal{H}}_0-\mu\,\hat{\mathcal{N}}_{\tm{el}})/T],
\end{equation}
where $Z=Z_{\tm{el}}Z_{\tm{ph}}$ is the grand partition function, $\hat{\mathcal{N}}_{\tm{el}}$ is the operator for the total number of electrons, and $\hat{\mathcal{H}}_0$ is the Hamiltonian of the uncoupled subsystems
\begin{equation}\label{Hep}
 \hat{\mathcal{H}}_0=\hat{H}_{\tm{el}}+\hat{H}_{\tm{ph}},
\end{equation}
whose eigenstates, constituing the thermodynamic basis, are
\begin{equation}\label{a0}
\ket{\bm{\alpha}}=\ket{n_1n_2\cdots n_k\cdots}\ket{N_1N_2\cdots N_q\cdots}=\ket{n;N}, 
\end{equation}
which represent the number of electrons, $\lbrace n_k\rbrace$, and phonons, $\lbrace N_q\rbrace$, in each single-particle state. 

The electric field is turned on at time $t=0^{+}$ to a constant value, i.e. $\bm{E}_t=\bm{E}$ for $t>0$, and the subsystems are subsequently coupled. In the notation of section \ref{masdS} we then have in the generalized thermodynamic description
\begin{equation}
 \hat{\mathcal{H}}=\hat{\mathcal{H}}_0+\hat{H}_{F},\hspace{0.5cm}\hat{V}=\hat{H}_{\tm{el-ph}}.
\end{equation}
Note that $\hat{V}$ is the coupling which fully mixes the degrees of freedom of the different subsystems (like the $\hat{H}_{lm}$ in section \ref{loceq}), which need not be separated by spatial boundaries. We now explain with some detail how the perturbation scheme developed in section \ref{masdS} applies to the present case. However, we only need to concentrate on how the transport equation is obtained in the Born-Markov limit, since this suffices to get the average rate of entropy production.

The idea is then to first derive the transport equation for the total system from the Liouville-von Neumann equation; we do it much in the same spirit as Kohn and Luttinger \cite{Kohn} did for elastic electronic scattering and generalized by Argyres \cite{Argyres} to inelastic scattering. Having this transport equation, the average rate of change of the total thermodynamic entropy in the Born-Markov limit is
\begin{equation}\label{dstr}
 \avg{\partial_t\hat{\mathcal{S}}_t}=-\sum_{\bm{\alpha}}(\partial_tP_{\bm{\alpha}})\ln P_{\bm{\alpha}},
\end{equation}
as can easily be verified by using \eqref{pau} in \eqref{bdS}. By proceeding with the transport or quantum Boltzmann equation for the electronic subsystem, we obtain a simple expression for the electronic entropy production. 

For the purpose of the present discussion, it suffices to work with the Liouville-von Neumann equation to first order in the electric field. That is, with $\hat{\rho}_t=\hat{\rho}_0+\hat{\rho}_{1;t}$ and $\hat{\rho}_{1;t}$ linear in the electric field, we write
\begin{equation}\label{r1}
i\partial_t \hat{\rho}_{1;t}=[\hat{\mathcal{H}}_0+\hat{V},\hat{\rho}_{1;t}]+[\hat{H}_F,\hat{\rho}_0], 
\end{equation}
where $\hat{\rho}_{1;0}=0$. The Laplace transform of this equation, with $\hat{\rho}_{1;s}=\int_0^\infty e^{-st}\hat{\rho}_{1;t}$, reads
\begin{equation}
is\hat{\rho}_{1;s}=(\mathcal{L}_0+L_V)\hat{\rho}_{1;s}+s^{-1} L_F\hat{\rho}_0. 
\end{equation}
With $\hat{\varrho}_{1;s}=\mathcal{D}\hat{\rho}_{1;s}$ and $\hat{\rho}_{1;s}^{\sim}=\mathcal{N}\hat{\rho}_{1;s}$, we separate this equation into a diagonal and a non-diagonal part, obtaining, respectively, the coupled algebraic equations
\begin{eqnarray}
is\hat{\varrho}_{1;s}&=&\mathcal{D}L_{V}\,\hat{\rho}_{1;s}^{\sim}+s^{-1}\mathcal{D}L_F\hat{\rho}_0,\label{d0}\\
\left[is+ \mathcal{N}(\mathcal{L}_0+L_V)\right]\hat{\rho}_{1;s}^{\sim}&=&\mathcal{N}L_{V}\,\hat{\varrho}_{1;s}+s^{-1}\mathcal{N}L_F\hat{\rho}_0.\hspace{0.5cm}\label{n0}
\end{eqnarray}
Solving for $\hat{\rho}_{1;s}^{\sim}$ in \eqref{n0} and substituting the result in \eqref{d0} we get a decoupled equation for $\hat{\varrho}_{1;s}$, which in the lowest Born approximation for the electron-phonon scattering reads
\begin{equation}
  is\hat{\varrho}_{1;s}=\mathcal{D}L_{V}\dfrac{1}{is+\mathcal{NL}_0}\mathcal{N}L_{V}\,\hat{\varrho}_{1;s}+s^{-1}\mathcal{D}L_F\hat{\rho}_0.
\end{equation}
From this, the transport equation for the total system easily arises in the Born-Markov limit by taking the Laplace inverse and neglecting memory terms. In terms of the occupation probabilities $P_{\bm{\alpha}}=\brket{\bm{\alpha}}{\hat{\varrho}_t}{\bm{\alpha}}$ we get 
\begin{equation}\label{ttot}
 \partial_tP_{\bm{\alpha}}=\dfrac{1}{i}(L_F\hat{\rho}_0)_{\bm{\alpha}}+\sum_{\bm{\alpha}'}(P_{\bm{\alpha}'}W_{\bm{\alpha}'\bm{\alpha}}-P_{\bm{\alpha}}W_{\bm{\alpha}\bm{\alpha}'})
\end{equation}
with the transition rates induced by the electron-phonon coupling $W_{\bm{\alpha}\bm{\alpha}'}=2\pi\delta(\varepsilon_{\bm{\alpha}}-\varepsilon_{\bm{\alpha}'})|\brket{\bm{\alpha}}{\hat{H}_{\tm{el-ph}}}{\bm{\alpha}'}|^2$. We have then derived the transport equation for the total system, in terms of which the average rate of change of the total thermodynamic entropy can be calculated, in the Born-Markov limit, using \eqref{dstr}.

To proceed with the calculation of the entropy production of the electronic subsystem, we note that
\begin{equation}\label{Pfa}
 P_{\bm{\alpha}}=P_n^{\tm{el}}P_{N}^{\tm{ph}}\chi_{nN}^{\tm{el-ph}},
\end{equation}
where $P_n^{\tm{el}}$ is the probability that the electrons are in the Fock state $\ket{n}$ regardless of the state of the phonons, $P_{N}^{\tm{ph}}$ is the probability that the phonons are in the Fock state $\ket{N}$ regardless of the state of the electrons, and $\chi_{nN}^{\tm{el-ph}}$ is the conditional probability that the total system is in the state  $\ket{\bm{\alpha}}$ in \eqref{a0}, \emph{given} that the electron and phonon subsystems are in states $\ket{n}$ and $\ket{N}$, respectively, without ``knowing'' about each other. Clearly, $\chi_{nN}^{\tm{el-ph}}$ is a function of the electron-phonon coupling strength, and can then be expanded in a power series of it
\begin{equation}
\chi_{nN}^{\tm{el-ph}}=1+\chi_{nN}^{\tm{el-ph}(1)}+\chi_{nN}^{\tm{el-ph}(2)}+\cdots.
\end{equation}
In the lowest Born approximation for the electron-phonon scattering, the electron and phonon subsystems are uncorrelated, i.e. $\chi_{nN}^{\tm{el-ph}}=1$, which is the usual Born-Oppenheimer approximation, and then by substituting \eqref{ttot} and \eqref{Pfa} into \eqref{dstr}, the average rate of change of the thermodynamic entropy of the total system turns out to be additive. For the electronic subsystem we have
\begin{equation}\label{dse}
 \avg{\partial_t\hat{\mathcal{S}}_t}_{\tm{el}}=-\sum_{n}(\partial_tP_{n}^{\tm{el}})\ln P_{n}^{\tm{el}},
\end{equation}
where the normalization condition $\sum_NP_{N}^{\tm{ph}}=1$ has been used. Here, the transport equation for the electronic subsystem is obtained from \eqref{ttot} by summing over $N$
\begin{equation}\label{dPn}
 \partial_tP_{n}^{\tm{el}}=\dfrac{1}{i}\sum_N(L_F\hat{\rho}_0)_{nN,nN}+\sum_{n'}(P_{n'}^{\tm{el}}\Gamma_{n'n}-P_{n}^{\tm{el}}\Gamma_{nn'}),
\end{equation}
where we have defined the phonon-averaged reduced transition rates $\Gamma_{nn'}$ as
\begin{equation}
\Gamma_{nn'}=\sum_N  P_{N}^{\tm{ph}}\sum_{N'}W_{nN,n'N'}.
\end{equation}
We can still go further and use the assumed statistical independence of the electrons to factorize their probability distribution into the probabilities of the one-electron states
\begin{equation}\label{Pnf}
P_{n}^{\tm{el}}=p_{n_1}p_{n_2}\cdots p_{n_k}\cdots,
\end{equation}
where $p_{n_k}$ is the probability that the one-electron state with quantum number $k$ has occupation $n_k=0,1$. Substituting this in \eqref{dse} we obtain an additive contribution to the average rate of change of the thermodynamic entropy of the electronic subsystem
\begin{equation}\label{dSel}
 \avg{\partial_t\hat{\mathcal{S}}_t}_{\tm{el}}=-\sum_{k,n_k}(\partial_tp_{n_k})\ln p_{n_k}=-\sum_k(\partial_tf_k)\ln\dfrac{f_k}{1-f_k},
\end{equation}
where in the last equality we identify the nonequilibrium one-electron distribution as $f_k=\sum_{n_k}n_kp_{n_k}=p_{n_k=1}$ and use  $\sum_{n_{k'}}p_{n_{k'}}=1$ to express $p_{n_k=0}=1-f_k$. The transport equation for $f_k$ is obtained by multiplying \eqref{dPn} by $n_k$ and summing over all $n$. To this end, note that
\begin{equation}\label{Gn}
\begin{split}
 \Gamma_{nn'}=\sum_{k,k'(k\neq k')}&w_{kk'}\,n_{k}(1-n_{k'})\\
 &\times|\overlap{\cdots n_{k'}-1\cdots n_{k}+1\cdots}{n'}|^2,
 \end{split}
\end{equation}
which is obtain by using \eqref{Helph} explicitly, where the one-electron transition rate from state $k$ to state $k^\prime$ is
\begin{eqnarray}
w_{kk'}&=&2\pi\sum_{q}\left|M_{k^\prime k}^q\right|^2\Big[\bar{N}(\omega_{q})\delta\left(\epsilon_{k^\prime}-\epsilon_k-\omega_{q}\right)
\nonumber
\\
&&+\left[1+\bar{N}(\omega_{q})\right]\delta\left(\epsilon_{k^\prime}-\epsilon_k+\omega_{q}\right)\Big],
\label{wdef}
\end{eqnarray} 
with $\bar{N}(\omega_q)=\sum_N P_N^{\tm{ph}}\brket{N}{\hat{a}_q^{\dagger}\hat{a}_q}{N}$ the average number of phonons in the single-particle state with quantum number $q$. We assume that the phonon subsystem can be kept in equilibrium at temperature $T$ (hence the dependence of $\bar{N}$ on $\omega_q$ only), no matter the nonequilibrium state of the electrons, as is the case for a good enough heat reservoir. 

That the phonons can be considered as a heat reservoir in the Born-Oppenheimer approximation can be seen by looking at the  transport equation for the phonon subsystem, obtained from \eqref{ttot} by summing over $n$
\begin{equation}\label{dPN}
 \partial_t P_N^{\tm{ph}}= \sum_{N'}(P_{N'}^{\tm{ph}}\Theta_{N'N}-P_{N}^{\tm{ph}}\Theta_{NN'}),
\end{equation}
where we have defined the electron-averaged reduced transition rates as
\begin{equation}
 \Theta_{NN'}=\sum_n P_n^{\tm{el}}\sum_{n'}W_{nN,n'N'}.
\end{equation}
Here we observe the important fact that the contribution from the first term of \eqref{ttot} vanishes due to the null value of the trace of the commutator $[\hat{H}_F,\hat{\rho}_0]$ in the subspace of electrons. This allows the existence of a steady state solution of \eqref{dPN} for which detailed balance holds, which is then an equilibrium solution. In any case, the assumption that the phonons are in equilibrium is not necessary for the following derivation of the electronic entropy production, as $\bar{N}(\omega_q)$ in \eqref{wdef} can be replaced  by the more complicated average obtained by using the nonequilibrium solution of \eqref{dPN}, not investigated here.

The transport equation for the one-electron distribution is then found to be, from \eqref{dPn}
\begin{equation}\label{Be1}
 \partial_tf_k=\dfrac{1}{i}\sum_{nN}n_k(L_F\hat{\rho}_0)_{nN,nN}+\sum_{nn'}n_k(P_{n'}^{\tm{el}}\Gamma_{n'n}-P_{n}^{\tm{el}}\Gamma_{nn'}).
\end{equation}
This is just the quantum Boltzmann equation. To write it in the familiar form we first note that, by writing $\hat{H}_{\tm{el}}$ in first-quantized form and using the well-known formula $[\hat{\bm{x}}_e,f(\hat{\bm{p}}_e)]=i\nabla_{\hat{\bm{p}}_e}f(\hat{\bm{p}}_e)$ we have $L_F\hat{\rho}_0=i(e/T)\bm{E}\cdot\hat{\bm{v}}\hat{\rho}_0$, where $\hat{\bm{v}}=\sum_e(\hat{\bm{p}}_e/m)=\sum_k \bm{v}_k \hat{c}_k^{\dagger}\hat{c}_k$ is the velocity operator of all electrons, with $\bm{v}_k=\nabla_k \epsilon_k$ the band velocity. Therefore, the first term in \eqref{Be1} is
\begin{equation}\label{dpd}
\begin{split}
 (\partial_t f_k)_{\textrm{drift}}&=\dfrac{1}{i}\sum_{nN}n_k(L_F\hat{\rho}_0)_{nN,nN}=\dfrac{e\bm{E}}{T}\cdot \tm{Tr}\,(\hat{n}_k\hat{\bm{v}}\hat{\rho}_0),\\
 &=e\bm{E}\cdot\bm{v}_k f_{k}^0(1-f_k^0)/T=-e\bm{E}\cdot\nabla_kf_k^0,
 \end{split}
\end{equation}
where $f_k^0=\tm{Tr}(\hat{n}_k\hat{\rho}_0)$ is the equilibrium Fermi-Dirac one-electron distribution. The second term in \eqref{Be1} can be written, using \eqref{Gn}, as
\begin{equation}
 \begin{split}
  (\partial_t f_k)_{\textrm{coll}}&=\sum_{nn'}n_k(P_{n'}^{\tm{el}}\Gamma_{n'n}-P_{n}^{\tm{el}}\Gamma_{nn'}),\\
  &=\sum_nP_n^{\tm{el}}n_k \sum_{k',k''}w_{k'k''}(1-n_{k'})n_{k''}\\
  &\ -\sum_nP_n^{\tm{el}}n_k \sum_{k',k''}w_{k'k''}n_{k'}(1-n_{k''}).
 \end{split}
\end{equation}
Therefore, by noting that $n_k(1-n_k)=0$ and using \eqref{Pnf} we see that the terms which do not cancel in the above sums are
\begin{equation}\label{dpc}
 (\partial_t f_k)_{\textrm{coll}}=\sum_{k'}[f_{k'}\,w_{k'k}\left(1-f_k\right)-f_{k}\,w_{kk'}\left(1-f_{k^\prime}\right)].
\end{equation}
We have thus arrived to the familiar form of the quantum Boltzmann equation by substituting \eqref{dpd} and \eqref{dpc} into \eqref{Be1}. With this, we can rewrite the average rate of change of the thermodynamic entropy of the electronic subsystem, from \eqref{dSel}, as
\begin{equation}\label{sebal}
 \avg{\partial_t\hat{\mathcal{S}}_t}_{\tm{el}}=\Pi_{\tm{el}}-\Phi_{\tm{el}},
\end{equation}
which is the entropy balance equation for the electronic subsystem, with the average electronic entropy production rate 
\begin{eqnarray}
\Pi_{\textrm{el}}&=&\dfrac{1}{2}\sum_{kk'}\left[f_{k'}\,w_{k'k}\left(1-f_k\right)-f_{k}\,w_{kk'}\left(1-f_k^\prime\right)\right]
\nonumber \\
&&\times\,\ln\dfrac{f_{k'}\,w_{k'k}\left(1-f_k\right)}{f_{k}\,w_{kk'}\left(1-f_k^\prime\right)},
\label{Piphonon}
\end{eqnarray}
which, similar to \eqref{Pi}, is a sum of terms of the form $(x-y)\ln(x/y)$ and then satisfies the second law of thermodynamics; and the entropy flux from the electrons to the phonons is
\begin{eqnarray}
\Phi_{\textrm{el}}&=&\dfrac{1}{2}\sum_{kk'}\left[f_{k'}\,w_{k'k}\left(1-f_k\right)-f_{k}\,w_{kk'}\left(1-f_k^\prime\right)\right]\ln\dfrac{w_{k'k}}{w_{kk'}}
\nonumber
\\
&&+\sum_k (\partial_t f_k)_{\textrm{drift}}\,\ln\dfrac{f_k}{1-f_k}.
\label{Phiphonon}
\end{eqnarray}
In the steady state the left-hand side of \eqref{sebal} is exactly zero and then all the entropy produced in the electronic system is transported to the phonons. We now want to show that this steady-state entropy flux toward the lattice vibrations gives the expression of the well-known Joule heating.

We need a solution, $f_k=f_k^0+\delta f_k$, of the quantum Boltzmann equation which, to linear order in the electric field strength, we write formally as
\begin{equation}
\delta f_k=\sum_{k'}\mathcal{W}_{kk'}^{-1}\left[\frac{e\bm{E}\cdot\bm{v}_{k'}}{T} f_{k'}^0\left(1-f_{k'}^0\right)\right],
\label{qbesollin}
\end{equation}
where the linearized collision operator $\mathcal{W}$, has matrix elements
\begin{equation}
\mathcal{W}_{kk'}=f_k^0w_{kk^\prime}+w_{k^\prime k}\left(1-f_k^0\right)-\frac{\delta_{kk^\prime}}{\tau_k}.
\label{Wdef}
\end{equation} 
The quasiparticle relaxation time $\tau_k$ is given by
\begin{equation}
\dfrac{1}{\tau_k}=\sum_{k'}\,[\,f_{k'}^0w_{k'k}+w_{kk'}(1-f_{k'}^0)\,],
\end{equation}
and becomes equal to the momentum relaxation time if the transition rates $w_{kk^\prime}$ are independent of the angle between $k$ and $k^\prime$.  

We now expand \eqref{Piphonon}. Because both the logarithm and the prefactor vanish in equilibrium the leading contribution is $O(\delta f)^2$. The term from expanding the logarithm is easily seen to be
\begin{equation}
\frac{\delta f_{k^\prime}}{f^0_{k^\prime}\left(1-f^0_{k^\prime}\right)}-\frac{\delta f_k}{f^0_k\left(1-f^0_k\right)},
\nonumber
\end{equation}
while the term coming from the prefactor is
\begin{equation}
\mathcal{W}_{kk^\prime}\delta f_{k'}-\mathcal{W}_{k^\prime k}\delta f_k.
\label{prefactor}
\end{equation}
Combining these equations with \eqref{qbesollin} yields
\begin{equation}
\Pi_{\textrm{el}}=\sum_k\frac{e\bm{E}\cdot\bm{v}_k}{T} \delta f_k=\frac{\sigma E^2}{T},
\label{Pifinal}
\end{equation}
where, in the last equality, we recognize the electric current as $e\avg{\hat{\bm{v}}}=\sum_ke\bm{v}_k\delta f_k=\sigma \bm{E}$ , with $\sigma$ the electric conductivity. Thus we see that to leading order in the electron-phonon coupling and the electric field, and on the assumption that the phonons act as a reservoir,  the electronic entropy production predicted by our formula is exactly the result expected from the Joule heating, $T\Pi_{\textrm{el}}=\sigma E^2$, implied by the electric field. Therefore, as desired, we have arrived at an expression of energy dissipation from a first-principle calculation of entropy production, not the other way around, as in previous approaches.

We remark that the results for the entropy production presented here are beyond the linear response theory. This is because, even when starting from the linear in the electric field correction to the density matrix, $\hat{\rho}_{1;t}$ (see Eq. \eqref{r1}), we derived the leading contribution to the electronic entropy production which is quadratic in the electric field. This is in contrast to past approaches \cite{Kohn,Suzuki2} for the calculation of the Joule heating, which requires going to the second order in the electric field contribution to the density matrix $\hat{\rho}_{2;t}$ for the calculation of the rate of change of the energy of the electrons. A field-theoretic approach \cite{Rammer1,Rammer2} for the calculation of higher order terms in the entropy production, beyond the Born-Markov approximation will be treated elsewhere. 

It is illustrative to evaluate the result explicitly, assuming e.g. dispersionless optical phonons $\omega_q = \omega_0$. With  $|M_{k'k}^q|^2=M\,\delta_{q,k'-k}$,  and assuming a degenerate electron system (i.e. $T\ll \epsilon_F$) we obtain
\begin{equation}\label{pie}
\Pi_{\textrm{el}}=\dfrac{(e^2E^2/3\pi m M)\,D_{\epsilon_F}}{D_{\epsilon_F-\omega_0}+D_{\epsilon_F+\omega_0}}(\epsilon_F/T)\sinh(\omega_0/T),
\end{equation}
with $D_\epsilon$ the electron density of states. In this case, the entropy production becomes large at low temperatures due to an increase in the conductivity (phonons not thermally activated and then scarcity of scattering centers), and hence in the Joule heating; this is expected when the only scattering mechanism is from optical phonons. 

Finally, we would like to point out the connection of the result \eqref{Pifinal} with the discussion in section \ref{loceq} concerning the foundations of the classical theory. With only the action of one of the subsystems (the sources of the $\bm{E}$-field) treated parametrically, with the three spatial components of the field $E_\lambda$ playing the role of the external parameters to the electronic subsystem, we can define an operator $\hat{F}_{\lambda}=\partial \hat{H}_F/\partial E_\lambda$ for the force exerted on the electrons upon variation of the field and write
\begin{equation}\label{Pil}
\begin{split}
T\hat{\Pi}_{\tm{el}}&=\sum_{\lambda} \hat{F}_{\lambda}\partial_t E_{\lambda}=\partial_t \sum_{\lambda} \hat{F}_{\lambda} E_{\lambda}-\sum_{\lambda} (\partial_t\hat{F}_{\lambda}) E_{\lambda}\\
&=\partial_t \hat{H}_F + e\hat{\bm{v}}\cdot \bm{E}=e\hat{\bm{v}}\cdot \bm{E},
\end{split}
\end{equation}
where $ \hat{H}_F=-e\sum_{\lambda,e} E_{\lambda}\hat{x}_e^{\lambda}$, and to get the last equality we use $\partial_t \hat{H}_F=\partial_t \hat{H}=0$, since the total system is isolated. Taking expectation value of \eqref{Pil}, the last equality is just \eqref{Pifinal} and the form of the first equality is reminiscent of the classical expression \eqref{Picl}. 

We then see that, although in the present discussion the subsystems are not separated by spatial boundaries (the essence of the generalized thermodynamic description) and there is no local equilibrium  at all times: the phonons remain in equilibrium, as implied by the assumption that they constitute a good heat reservoir, but the electrons attain a nonequilibrium steady state; the common feature with the discussion in section \ref{loceq} is the complete factorization of the probability distribution of the system over the degrees of freedom of the different subsystems (uncorrelated subsystems), here manifested as the Born-Oppenheimer approximation. An appropriate account of the quantum correlations between subsystems is therefore the key to purely quantum thermodynamic behavior.

\section{Conclusion}\label{con}
We have developed a  theory for the entropy production in quantum many-body systems by introducing an entropy operator and calculating the average rate of change of its thermodynamically measurable part. We show that the laws of thermodynamics are satisfied exactly within our formalism. In the Born-Markov approximation which describes the physics of weakly-coupled subsystems of an isolated system in the long-time limit, the theory reproduces the entropy balance equation which is fundamental in classical nonequilibrium thermodynamics and the Joule heating contribution to the entropy production expected in a standard conductor. Applications to other systems as well as generalizations beyond the weak-coupling limit will be presented elsewhere.  

\section*{acknowledgments}
ESC would like to acknowledge the support from the Fulbright-Colciencias fellowship. The work of AJM was supported by the National Science Foundation under grant DMR-1308236.
\bibliography{references}

\end{document}